\documentclass{article}
\usepackage{amsthm, amsmath, amsfonts, amssymb, charter}
\usepackage{authordate1-4}
\usepackage{tikz}
\usetikzlibrary{arrows}
\begin{document}

\renewcommand{\qedsymbol}{$\blacksquare$}
\newtheorem{thm}{Theorem} [section]     
\newtheorem{cor}[thm]{Corollary} 
\newtheorem{lem}[thm]{Lemma} 
\newtheorem{prop}[thm]{Proposition} 
\newtheorem{rem}[thm]{Remark}

\theoremstyle{definition}
\newtheorem{dfn}[thm]{Definition}
\newtheorem{ex}[thm]{Example}

\title{Some Non-Classical Approaches to the Branderburger-Keisler Paradox}
\author{\textbf{Can Ba\c{s}kent}}
\date{\small{Department of Computer Science, The Graduate Center \\ The City University of New York, New York, USA. \\ \textsf{cbaskent@gc.cuny.edu}, \textsf{www.canbaskent.net}}}
\maketitle
\begin{abstract}
In this paper, we discuss a well-known self-referential paradox in foundational game theory, the Brandenburger - Keisler paradox. We approach the paradox from two different perspectives: non-well-founded set theory and paraconsistent logic. We show that the paradox persists in both frameworks for category theoretical reasons, but, with different properties.
\end{abstract}

\section{Introduction}

\subsection{Motivation} 
The Brandenburg-Keisler paradox (`BK paradox', henceforth) is a two-person self-referential paradox in epistemic game theory \cite{bran0}. Due to its considerable impact on the various branches of game theory and logic, it has gained increasing interest in the literature.

In short, for players Ann and Bob, the BK paradox arises when we consider the following statement  ``Ann believes that Bob assumes that Ann believes that Bob's assumption is wrong" and ask the question if ``Ann believes that Bob's assumption is wrong".

There can be considered two main reasons why the Brandenburger-Keisler argument turns out to be a  paradox. First, the limitations of set theory presents some restrictions on the mathematical model which is used to describe self-referantiality and circularity in the formal language. Second, Boolean logic comes with its own Aristotelian meta-logical assumptions about consistency. Namely, Aristotelian principle about consistency, \emph{principium contradictionis}, maintains that contradictions are impossible. In this paper, we will consider some alternatives to such assumptions, and investigate their impact on the BK paradox.

The BK paradox is based on the ZFC set theory. The ZFC set theory comes with its own \emph{restrictions} one of which is the \emph{axiom of foundation}. It can be deduced from this axiom that no set can be an element of itself. In non-well-founded set theory, on the other hand, the axiom of foundation is replaced by the \emph{anti-foundation axiom} which leads to, among many other things, generation of sets which are members of themselves \cite{mir,acz}. Therefore, we claim that switching to non-well-founded set theory suggests a new approach to the paradox, and game theory in general. The power of non-well-founded set theory comes from its genuine methods to deal with circularity \cite{bar1,moss0}.

Second, what makes the BK paradox a \emph{paradox} is the \emph{principium contradictionis}. Paraconsistent logics challenge this assumption \cite{pri1,pri}. Therefore, we also investigate the BK paradox in paraconsistent systems. This line of research, as we shall see, is rather fruitful.  The reason for this is the following. The BK paradox is essentially a self-referential paradox, and similar to any other paradox of the same kind, it can be analyzed from a category theoretical or algebraic point of view \cite{yan,abr0}. Moreover, paraconsistent logics also present an algebraic and category theoretical structure which makes this approach possible. In this work, we make the connection between self-referantiality and paraconsistency clearer and see whether we can \emph{solve} the paradox if we embrace a paraconsistent framework.

What is the significance of adopting non-classical frameworks then? There are many situations where circularity and inconsistency are integral parts of the game. A game when some players can \emph{reset} the game can be thought of a situation where the phenomenon of circularity appears. Moreover, inconsistencies occur in games quite often as well. Situations where information sets of some players become inconsistent after receiving some information in a dialogue without a consequent belief revision are such examples where inconsistencies occur \cite{leb,rah}. 

In this paper, we show that adopting the non-well-founded set theory makes a significant change in the structure of the paradox. We achieve this by constructing counter-models for the BK argument. Second, by paraconsistent logic, we show that, even when we allow non-trivial inconsistencies the BK argument \emph{can} be satisfied in some certain situations. We also use topological products to give a weak-completeness result generalizing some of the results in the original paper.

\subsection{Related Literature} 

The Brandenburger - Keisler paradox was presented in its final form in a relatively recent paper in 2006 \cite{bran0}. This paper was followed by some further result within the same domain \cite{bran1}. 

A general framework for self-referential paradoxes was discussed earlier by Yanofsky in 2003 \cite{yan}. In his paper, Yanofsky used Lawvere's category theoretical arguments in well-known mathematical arguments such as Cantor's diagonalization, Russell's paradox, and G\"{o}del's Incompleteness theorems. Lawvere, on the other hand, discussed self-referential paradoxes in cartesian closed categories in his early paper which appeared in 1969 \cite{law0}. Most recently, Abramsky and Zvesper used Lawvere's arguments to analyze the BK paradox in a category theoretical framework \cite{abr0}.

Pacuit approached the paradox from a modal logical perspective and presented a detailed investigation of the paradox in neighborhood models and in hybrid systems \cite{pac2}. Neighborhood models are used to represent modal logics weaker than $\mathbf{K}$, and can be considered as weak versions of topological semantics \cite{che}. This argument later was extended to assumption-incompleteness in modal logics \cite{pac4}.

To the best of our knowledge, the idea of using non-well-founded sets as Harsanyi type spaces was first suggested by Lismont, and extended later by Heifetz \cite{lis,heif}. Heifetz motivated his approach by ``making the types an explicit part of the states' structure", and hence obtained a circularity that enabled him to use non-well-founded sets.

Mariotti et al., on the other hand, used compact belief models to represent interactive belief structures in a topological framework with further topological restrictions \cite{mar}.

Paraconsistent games in the form of dialogical games were largely discussed by Rahman and his co-authors \cite{rah}. Co-Heyting algebras, on the other hand, has gained interest due to their use in ``region based theories of space" within the field of \emph{mereotopology} \cite{ste2}. Mereotopology discusses the qualitative topological relations between the wholes, parts, contacts and boundaries and so on.

The organization of the paper is as follows. First, we recall the BK paradox stated in basic modal language. Then, we consider the matter from non-well-founded set theoretical point of view. Next, we introduce category theoretical and topological paraconsistent frameworks to deal with the paradox, and analyze the behavior of the paradox in such frameworks. Finally, we conclude with several research directions for future work.

\subsection{The Paradox}

The BK paradox can be considered as a game theoretical two-person version of Russell's paradox where players interact in a self-referential fashion. Let us call the players Ann and Bob with associated type spaces $U^{a}$ and $U^{b}$ respectively. Now, consider the following statement which we call the \emph{BK sentence}:
\begin{quote}
\emph{Ann believes that Bob assumes that Ann believes that Bob's assumption is wrong.}
\end{quote}

A Russell-like paradox arises if one asks the question whether \emph{Ann believes that Bob's assumption is wrong}. In both cases, we get a contradiction, hence the paradox. Thus, the BK sentence is impossible.

Brandenburger and Keisler use belief sets to represent the players' beliefs. The model $(U^{a}, U^{b}, R^{a}, R^{b})$ that they consider is called a \emph{belief model}  where $R^{a} \subseteq U^{a} \times U^{b}$ and $R^{b} \subseteq U^{b} \times U^{a}$. The expression $R^{a}(x, y)$ represents that in state $x$, Ann believes that the state $y$ is possible for Bob, and similarly for $R^{b}(y, x)$. We will put $R^{a}(x) = \{ y : R^{a}(x, y) \}$, and similarly for $R^b(y)$. At a state $x$, we say Ann believes $P \subseteq U^{b}$ if $R^{a}(x) \subseteq P$. Now, a modal logical semantics for the interactive belief structures can be given. We use two different modalities $\Box$ and $\heartsuit$ which stand for the belief and assumption operators respectively with the following semantics.

\begin{quote}\begin{tabular}{lll}
$x \models \Box^{ab} \varphi $ & iff & $\forall y \in U^{b}. R^{a}(x, y)$ implies $y \models \varphi$ \\
$x \models \heartsuit^{ab} \varphi $ & iff & $\forall y \in U^{b}. R^{a}(x, y)$ iff $y \models \varphi$ \\
\end{tabular}\end{quote}

Those operators can also be given a modal definition \cite{bran0}. An interactive belief frame is a structure $(W, P, U^a, U^b)$ with a binary relation $P \subseteq W \times W$, and disjoint sets $U^a$ and $U^b$ such that $(U^a, U^b, P^a, P^b)$ is a belief model with $U^a \cup U^b = W$, $P^a = P \cap U^a \times U^b$, and $P^b = P \cap U^b \times U^a$. Now, for a given valuation function which assigns propositional variables to subsets of $W$, the semantics of the belief and assumption modalities are given as follows.

\begin{quote}\begin{tabular}{lll}
$x \models \Box^{ab} \varphi $ & iff & $w \models \mathbf{U}^a \wedge \forall y (P(x, y) \wedge y \models \mathbf{U}^b$ implies  $y \models \varphi)$ \\
$x \models \heartsuit^{ab} \varphi $ & iff & $w \models \mathbf{U}^a \wedge \forall y (P(x, y) \wedge y \models \mathbf{U}^b$ iff  $y \models \varphi)$ \\
\end{tabular}\end{quote}

A belief structure $(U^{a}, U^{b}, R^{a}, R^{b})$ is called \emph{assumption complete} with respect to a set of predicates $\Pi$ on $U^{a}$ and $U^{b}$ if for every predicate $P \in \Pi$ on $U^{b}$, there is a state $x \in U^{a}$ such that $x$ assumes $P$, and for every predicate $Q \in \Pi$ on $U^a$, there is a state $y \in U^b$ such that $y$ assumes $Q$. We will use special propositions $\mathbf{U}^a$ and $\mathbf{U}^b$ with the following meaning: $w \models \mathbf{U}^a$ if $w \in U^a$, and similarly for  $\mathbf{U}^b$. Namely, $\mathbf{U}^a$ is true at each state for player Ann, and $\mathbf{U}^b$ for player Bob.

Brandenburger and Keisler showed that no belief model is complete for its first-order language. Therefore, ``not every description of belief can be represented" with belief structures \cite{bran0}. The incompleteness of the belief structures is due to the \emph{holes} in the model. A model, then, has a hole at $\varphi$ if either $\mathbf{U}^b \wedge \varphi$ is satisfiable but $\heartsuit^{ab} \varphi$ is not, or $\mathbf{U}^a \wedge \varphi$ is satisfiable but $\heartsuit^{ba} \varphi$ is not. A big hole is then defined by using the belief modality $\Box$ instead of the assumption modality $\heartsuit$.

In the original paper, the authors make use of two lemmas before identifying the holes in the system. These lemmas are important for us as we will challenge them in the next section. First, let us define a special propositional symbol $\mathbf{D}$ with the following  valuation $D = \{ w \in W : (\forall z \in W) [P(w, z) \rightarrow \neg P(z, w)] \}$.

\begin{lem}[\cite{bran0}]{\label{lemma-bk}} ~
\begin{enumerate}
\item If $\heartsuit^{ab} \mathbf{U}^b$ is satisfiable, then $\Box^{ab} \Box^{ba} \Box^{ab} \heartsuit^{ba} \mathbf{U}^a \rightarrow \mathbf{D}$ is valid.
\item $\neg \Box^{ab} \heartsuit^{ba} (\mathbf{U}^a \wedge \mathbf{D})$ is valid.
\end{enumerate} \end{lem}

Based on these lemmas, authors observe that there is no complete belief models. Here, we give the theorem in two forms.

\begin{thm}[\cite{bran0}] ~{\label{bk-thm}}
\begin{itemize}
\item First-Order Version: Every belief model $M$ has either a hole at $U^a$, a hole at $U^b$, a big hole at one of the formulas
\begin{itemize}
\item[(i)] $\forall x. P^b(y, x)$
\item[(ii)] $x$ believes $\forall x. P^b(y, x)$
\item[(iii)] $y$ believes [$x$ believes $\forall x. P^b(y, x)$]
\end{itemize}
a hole at the formula
\begin{itemize}
\item[(iv)] $D(x)$
\end{itemize}
or a big hole at the formula
\begin{itemize}
\item[(v)] $y$ assumes $D(x)$
\end{itemize}
Thus, there is no belief model which is complete for a language $\mathcal{L}$ which contains the tautologically true formulas and formulas (i)-(v).

\item Modal Version: There is either a hole at $\mathbf{U}^a$, a hole at $\mathbf{U}^b$, a big hole at one of the formulas $$\heartsuit^{ba}\mathbf{U}^a, \qquad  \Box^{ab} \heartsuit^{ba}\mathbf{U}^a, \qquad  \Box^{ba} \Box^{ab} \heartsuit^{ba}\mathbf{U}^a$$
a hole at the formula $\mathbf{U}^a \wedge \mathbf{D}$, or a big hole at the formula $\heartsuit^{ba}(\mathbf{U}^a \wedge \mathbf{D})$. Thus, there is no complete interactive frame for the set of all modal formulas built from $\mathbf{U}^a$, $\mathbf{U}^b$, and $\mathbf{D}$.
\end{itemize} \end{thm}

\section{Non-well-founded Set Theoretical Approach}

\subsection{Introduction}

Non-well-founded set theory is a theory of sets where the axiom of foundation is replaced by the \emph{anti-foundation axiom} which is due to Mirimanoff \cite{mir}. Decades later, the axiom was re-formulated by Aczel within the domain of graph theory which motivates our approach here \cite{acz}. In non-well-founded (NWF, henceforth) set theory, we can have true statements such as `$x \in x$', and such statements present interesting properties in game theory. NWF theories, in this respect, are natural candidates to represent circularity \cite{bar1}. 

To the best of our knowledge, Lismont introduced non-well-founded type spaces to show the existence of universal belief spaces \cite{lis}. Then, Heifetz used NWF sets to represent type spaces and obtained  rather sophisticated results \cite{heif}. He mapped a given belief space to its NWF version, and then proved that in the NWF version, epimorphisms become equalites.

However, Harsanyi noted earlier that circularity might be needed to express infinite hierarchy of beliefs.

\begin{quote}
It seems to me that the basic reason why the theory of games with incomplete information has made so little progress so far lies in the fact that these games give rise, or at least appear to give rise, to an infinite regress in reciprocal expectations on the part of the players. In such a game player 1's strategy choice will depend on what he expects (or believes) to be player 2's payoff function $U_2$, as the latter will be an important determinant of player 2's behavior in the game. But his strategy choice will also depend on what he expects to be player 2's first-order expectation about his own payoff function $U_1$. Indeed player 1's strategy choice will also depend on what he expects to be player 2's second-order expectation - that is, on what player 1 thinks that player 2 thinks that player 1 thinks about player 2's payoff function $U_2$... and so on \emph{ad infinitum}.”

\cite{hars}
\end{quote}

Note that Harsanyi's concern for infinite regress or circularity is related to the epistemics of the game. However, some other ontological concerns can be raised as well about the type spaces, and the way we define the states in the type spaces. In this respect, Heifetz motivated his approach, which is related to our perspective here, by arguing that NWF type spaces can be used ``once states of nature and types would no longer be associated with states of the world, but constitute \emph{their very definition}." [ibid, (his emphasis)]. 

This is, indeed, a prolific approach to Harsanyi type spaces to represent uncertainty. Here is Heifetz on the very same issue.
\begin{quote}
Nevertheless, one may continue to argue that a state of the world should indeed be a circular, self-referantial object: A state represents a situation of human uncertainty, in which a player considers what other players may think in other situations, and in particular about what they may think there about the current situation. According to such a view, one would seek a formulation where states of the world are indeed self-referring mathematical entities.

\cite[p. 204]{heif}.
\end{quote}

Notice that BK paradox is a situation where the aforementioned belief interaction among the players plays a central role. Therefore, in our opinion, it is worthwhile to pursue what NWF type spaces might provide in such situations. On the other hand, NWF set theory is not immune to the problems that the classical set theory suffers from. For example, note that Russell's paradox is not solved in NWF setting, and moreover the subset relation stays the same in NWF theory \cite{moss0}. The reason is quite straight-forward. As Heifetz also noted ``Russell's paradox applies to the collection of all sets which do not contain themselves, not to the collection of sets which \emph{do} contain themselves" \cite[(his emphasis)]{heif}. Therefore, we do not expect the BK paradox to disappear in NWF setting. Yet, NWF set theory will give us many other tools in game theory. We will also go back to this issue when we use category theoretical tools.

\subsection{Non-well-founded Belief Models}

Let us start with defining belief models in NWF theory. What we call a non-well-founded model is a tuple $M = (W, V)$ where $W$ is a non-empty non-well-founded set (\emph{hyperset}, for short), and $V$ is a valuation assigning propositional variables to the elements of $W$.

Now, we give the semantics of (basic) modal logic in non-well-founded setting \cite{ger}. We use the symbol $\models^+$ to represent the semantical consequence relation in a NWF model.

\begin{quote}\begin{tabular}{lll}
$M, w \models^+ \Diamond \varphi$ & iff & $\exists v \in w.$ such that $M, v \models^+ \varphi$ \\
$M, w \models^+ \Box \varphi$ & iff & $\forall v \in w$. $v \in w$ implies $M, v \models^+ \varphi$
\end{tabular}\end{quote}\

Based on this definition, we can now give a non-standard semantics for the belief and assumption modalities $\Box^{ij}$ and $\heartsuit^{ij}$ respectively for $i, j \in \{ a, b \}$.

\begin{quote}\begin{tabular}{lll}
$M, w \models^+ \Box^{ij} \varphi$ & iff & $M, w \models^+ \mathbf{U}^i ~\wedge$ \\ && $\forall v (v \in w \wedge M, v \models^+ \mathbf{U}^j \rightarrow M, v \models^+ \varphi)$\\
$M, w \models^+ \heartsuit^{ij} \varphi$ & iff & $M, w \models^+ \mathbf{U}^i ~\wedge$ \\ && $\forall v (v \in w \wedge M, v \models^+ \mathbf{U}^j \leftrightarrow M, v \models^+ \varphi)$
\end{tabular}\end{quote}

Several comments on the NWF semantics are in order here. First, notice that this definition of NWF semantics for belief and assumption modalities depend on the earlier modal definition of those operators given \cite{bran0,ger}. Second, belief or assumption of $\varphi$ at a state $w$ is defined in terms of the truth of $\varphi$ at the states that constitutes $w$, including possibly $w$ itself. Therefore, these definitions address the philosophical and foundational points that Heifetz made about the uncertainty in type spaces. We call a belief state $w \in W$ \emph{Quine state} if $w = \{ w \}$. We call a belief state $w \in W$ an \emph{urelement} if it is not the empty set, and it can be a member of a set but cannot have members. Finally, we call a set $A$ \emph{transitive} if $a \in A$ and $b \in a$, then $b \in A$.

For example consider the model $W = \{ w, v\}$ with $V(p) = \{ w \}$ and $V(q) = \{ v \}$ with a language with two propositional variables for simplicity. Let us assume that both $w$ and $v$ are Quine states. What does it mean to say that the player $a$ assumes $p$ at $w$ in NFW belief models? Since $w$ is a Quine state, the only state it can access is itself. Therefore, the statement $w \models^+ \heartsuit^{ab} p$ forces $w \in U^a \cap U^b$, which is impossible since the type spaces of $a$ and $b$ are assumed to be disjoint. On the other hand, for $w \in U^a$, we have $w \models^+ \heartsuit^{ab} q$. Notice that $w \models^+ \mathbf{U}^a$. Moreover, since $w$ is the only member of $w$, and $w \not\models^+ \mathbf{U}^b$, together with the assumption that $w \not\models^+ q$, we observe that the bicondition is satisfied. Therefore, Quine states can only assume false statements. Moreover, they believe in any statements.

\begin{thm}
Let $M = (W, V)$ be a NWF belief model with disjoint type spaces $U^a$ and $U^b$ respectively for two players $a$ and $b$. If $w \in U^i$ be a Quine state or an urelement belief state for $i \in \{a,b\}$, then $i$ assumes $\varphi$ at $w$ if and only if $M, w \not\models^+ \varphi$. Moreover, $i$ believes in any formula $\psi$ at $w$.
\end{thm}

\begin{proof}
Let us first start with considering the Quine states. 

Without loss of generality, let $w \in U^a$ where $w$ is a Quine state. Suppose $w \models^+ \heartsuit^{ab} \varphi$. Since $w \in U^a$, $w \models^+ \mathbf{U}^a$. Since, $w \in w$, and $U^a$ and $U^b$ are disjoint, we have $w \not\models^+ \mathbf{U}^b$. Therefore, since  $w \models^+ \heartsuit^{ab} \varphi$, we conclude $w \not\models^+ \varphi$.

For the converse direction, suppose that $w \not\models^+ \varphi$. Since $w \in w$, and $w \not\models^+ \mathbf{U}^b$ the biconditional is satisfied. Moreover, since $w \in U^a$, $w \models+ \mathbf{U}^a$. Therefore, $w \models^+ \heartsuit^{ab} \varphi$.

The proof for the belief operator is immediate for Quine states.

Now, let us consider urelements. Without loss of generality, let $w \in U^a$ be an urelement. Then, $w \models^+ \mathbf{U}^a$. For the left-to-right direction, note that since there is no $v \in w$, the conditional is vacuously satisfied. For the right-to-left direction, suppose $w \not\models^+ \varphi$. Since, $w \notin w$ and $w \not\models^+ \mathbf{U}^b$, the biconditional is satisfied again.

The proof for the belief operator is also immediate for urelement belief states, and thus left to the reader.

This concludes the proof.
\end{proof}

Notice that the proof heavily depends on the fact that the type spaces for the players are assumed to be disjoint. Let us now see how belief models change once we allow the intersection of NWF type spaces.

\begin{thm}{\label{quine}}
Let $M = (W, V)$ be a NWF belief model for two players $a$ and $b$ where $U^a$ and $U^b$ is not necessarily disjoint. For a Quine state $w$, if $w \models^+ \heartsuit^{ij} \top$ for $i, j \in \{ a, b\}$, then $w \in U^a \cap U^b$. In other words, Quine states with true assumptions belong to the both players.
\end{thm}

\begin{proof}
Without loss of generality, assume that Quine state $w$ is in $U^a$. Suppose $w \models^+ \heartsuit^{ab} \top$. Then, we observe $w \models^+ \mathbf{U}^a$. Left-to-right direction of the biconditional is immediate. Now, consider the right-to-left direction. Since, $\top$ is satisfied at any state, and $w$ is non-empty (i.e. $w \in w$), we conclude that $w \models^+ \mathbf{U}^b$. Therefore, $w \in U^b$. Thus, $w \in U^a \cap U^b$.
\end{proof}

Game theoretical implications of Theorem~\ref{quine} is worth mentioning. From the standard belief structure's point of view, Quine states correspond to the states which are reflexive. In other words, at such a state $w$, player $i$ considers $w$ possible for player $j$. Thus, such a state $w$ is forced to be in the intersection of the type spaces.

On the other hand, intersecting type spaces do not seem to create a problem for belief models. To overcome this issue, one can introduce a \emph{turn} function from the space of the belief model to the set of players assigning states to players. The functional definition of this construction necessitates that every state should be assigned to a unique player. Therefore, the game can determine whose turn it is at Quine atoms. Additionally, urelements, since they cannot have elements, are end states in games. At such states, players do not consider any states possible for the other player.

Now, based on the NWF semantics we gave earlier, it is not difficult to see that the following formulas discussed in the original paper are still valid as before if we maintain the assumption of the disjointness of type spaces.

$$\Box^{ab} \mathbf{U}^b \leftrightarrow \mathbf{U}^a,~~ \Box^{ba} \mathbf{U}^a \leftrightarrow \mathbf{U}^b,~~ \Box^{ab} \mathbf{U}^a \leftrightarrow \bot,~~ \Box^{ba} \mathbf{U}^b \leftrightarrow \bot$$

Furthermore, the following formulas are not valid as before.

$$\Box^{ab} \mathbf{U}^b \rightarrow \mathbf{U}^b,~~ \Box^{ab} \mathbf{U}^b \rightarrow \Box^{ba} \Box^{ab} \mathbf{U}^b,~~ \Box^{ab} \mathbf{U}^b \rightarrow \Box^{ab} \Box^{ab} \mathbf{U}^b$$

However, for the sake of the completeness of our arguments, let us, for the moment, allow that type spaces may not be disjoint.

Consider a NWF belief model $(W, V)$ where $w = \{ w \}$ with $U^a = U^b = W$. In such a model $\Box^{ab}\mathbf{U}^a \leftrightarrow \bot$ fails, but    $\Box^{ab}\mathbf{U}^a \leftrightarrow \top$ is satisfied. Similar observations can be made for $\Box^{ba}\mathbf{U}^b \leftrightarrow \bot$ and $\Box^{ba}\mathbf{U}^b \leftrightarrow \top$. Similarly, all $\Box^{ab} \mathbf{U}^b \rightarrow \mathbf{U}^b$, $\Box^{ab} \mathbf{U}^b \rightarrow \Box^{ba} \Box^{ab} \mathbf{U}^b$, and $\Box^{ab} \mathbf{U}^b \rightarrow \Box^{ab} \Box^{ab} \mathbf{U}^b$ are satisfied in the aforementioned NWF model. 

Thus, following Heifetz's arguments, we maintain that NWF models with self-referring states are better candidates to formalize uncertainty in games. Therefore, based on the above observations, it is now conceivable to imagine a NWF belief model in which previously constructed standard holes do not exist. Our aim now is to construct a NWF belief model in which the Lemma~\ref{lemma-bk} fails. For this purpose of us, however, we still maintain the assumption that type space be disjoint.

As a first step, we redefine the diagonal set in the NWF setting. Recall that, in the standard case, diagonal set $D$ is defined with respect to the accessibility relation $P$ which we defined earlier.  In NWF case, we will use membership relation for that purpose.

\begin{dfn}
Define $D^+ := \{ w \in W : \forall v \in W . (v \in w \rightarrow w \notin v) \}$. 
\end{dfn}

We define the propositional variable $\mathbf{D}^+$ as the propositional variable with the valuation set $D^+$. 

Now, we observe how the NWF models make a difference in the context of the BK paradox. Notice that BK argument relies on two lemmas which we have mentioned earlier in Lemma~\ref{lemma-bk}. Now, we present counter-models to Lemma~\ref{lemma-bk} in NWF theory.

\begin{prop}
In a NWF belief structure, if $\heartsuit^{ab} \mathbf{U}^b$ is satisfiable, then the formula $\Box^{ab} \Box^{ba} \Box^{ab} \heartsuit^{ba} \mathbf{U}^a \wedge  \neg \mathbf{D}^{+}$ is also satisfiable.
\end{prop}

\begin{proof}
Let $W = \{ w, v \}$ with $w = \{ v \}$, $v = \{ w \}$ where $U^a = \{ w \}$ and $U^b = \{ v \}$. To maintain the disjointness of the types, assume that neither $w$ nor $v$ is transitive.

Then, $w \models^+ \heartsuit^{ab} \mathbf{U}^b$ since all states in $b$'s type space is assumed by $a$ at $w$. Similarly, $v \models^+ \heartsuit^{ba} \mathbf{U}^a$ as all states in $a$'s type space is assumed by $b$ at $v$. Then, $w \models^+ \Box^{ab} \heartsuit^{ba} \mathbf{U}^a$. Continuing this way, we conclude, $w \models^+ \Box^{ab} \Box^{ba} \Box^{ab} \heartsuit^{ba} \mathbf{U}^a$. 

However, by design, $w \not\models^+ \mathbf{D}^{+}$ since $v \in w$ and $w \in v$. Thus, the formula $\Box^{ab} \Box^{ba} \Box^{ab} \heartsuit^{ba} \mathbf{U}^a \wedge  \neg \mathbf{D}^{+}$ is satisfiable as well.
\end{proof}

\begin{prop}
The formula $\Box^{ab} \heartsuit^{ba} (\mathbf{U}^a \wedge \mathbf{D}^{+})$ is satisfiable in some NWF belief structures.
\end{prop}

\begin{proof}
Take a non-transitive model $(W, V)$ with $W = \{ w, v, u, t\}$ where $w = \{ v, w \}$, $v = \{ u \}$, and $u = \{ t \}$ where $u \notin t$. Let $U^a = \{ w, u\}$, and $U^b = \{ v, t \}$. Now, observe that the formula $\mathbf{U}^a \wedge \mathbf{D}^+$ is satisfiable only at $u$ (as $w \in w$, $w$ does not satisfy $\mathbf{U}^a \wedge \mathbf{D}^+$). Now, $v \models^+ \heartsuit^{ba} (\mathbf{U}^a \wedge \mathbf{D}^+)$. Finally,  $w \models^+ \Box^{ab} \heartsuit^{ba} \mathbf{U}^a \wedge \mathbf{D}^+$. Note that even if $w \in w$ as $w \notin U^b$, by definition of the box modality, $w$ satisfies $\Box^{ab} \heartsuit^{ba} \mathbf{U}^a \wedge \mathbf{D}^+$.
\end{proof}

Therefore, the Lemma~\ref{lemma-bk} is refuted in NWF belief models. Notice that Lemma~\ref{lemma-bk} is central in Brandenburger and Keisler's proof of the incompleteness of belief structures. Thus, we ask whether the failure of Lemma~\ref{lemma-bk} would mean that there can be complete NWF belief structures. The answer to this question requires some category theoretical tools that we will introduce in the following section. For now, we construct a counter-model for Theorem~\ref{bk-thm} in NWF setting.

Consider the following counter-model. Let $W = \{ w, v, u, t, r, s, x, y, z \}$ with $w = \{ v, w \}$, $v = \{ u \}$, $u = \{ t \}$ ($u \notin t$), $r = \{ v, t, s, y \}$, $s = \{ w, u, r, x, z \}$, $x = \{ x, s\}$, $y = \{ y, x \}$, $z = \{ z, y \}$ where $U^a = \{ w, u, r, x, z \}$ and $U^b = \{ v, t, s, y \}$. Now, observe the following.

\begin{itemize}
\item $\mathbf{U} \wedge \mathbf{D}^+$ is satisfied only at $u$, since we have $w \in w$, $r \in s \wedge s\in r$, $x \in x$ and $z \in z$
\item No hole at $\mathbf{U}^a$ as $s \models^+ \heartsuit^{ba} \mathbf{U}^a$
\item No hole at $\mathbf{U}^b$ as $r \models^+ \heartsuit^{ab} \mathbf{U}^b$
\item No big hole at $\heartsuit^{ba} \mathbf{U}^a$ as $x \models^+ \Box^{ab} \heartsuit^{ba} \mathbf{U}^a$
\item No big hole at $\Box^{ab} \heartsuit^{ba} \mathbf{U}^a$ as $y \models^+ \Box^{ba} \Box^{ab} \heartsuit^{ba} \mathbf{U}^a$
\item No big hole at $\Box^{ba} \Box^{ab} \heartsuit^{ba} \mathbf{U}^a$ as $z \models^+ \Box^{ab} \Box^{ba} \Box^{ab} \heartsuit^{ba} \mathbf{U}^a$
\item No hole at $\mathbf{U}^a \wedge \mathbf{D}^+$ as $v \models^+ \heartsuit^{ba} (\mathbf{U}^a \wedge \mathbf{D}^+)$
\item No big hole at $\heartsuit^{ba} (\mathbf{U}^a \wedge \mathbf{D}^+)$ as $w \models^+ \Box^{ab} \heartsuit^{ba} (\mathbf{U}^a \wedge \mathbf{D}^+)$
\end{itemize}

The crucial point in the semantical evaluation of big holes is the fact that the antecedent of the conditional in the definition of the box modality is not satisfied if some elements of the current states are not in the desired type space. Therefore, the entire statement of the semantics of the box modality is still satisfied if the current state has some elements from the same type space. This helped us to construct the counter-model.

This counter-model shows that Theorem~\ref{bk-thm} with its stated form does not hold in NWF belief structures. Yet, we have to be careful here. Our counter model does not establish the fact that NWF belief models are complete. It does establish the fact that they do not have the same holes as the standard belief models. We will get back to this question later on, and give an answer from category theoretical point of view.

Similarly, NWF theory should not be taken as such a revolutionary approach to epistemic game theory replacing the classical (Kripke) models. Gerbrandy noted:

\begin{quote}
(...)[T]here are many `more' Kripke models than there are possibilities of knowledge structures: each possibly corresponds [to] a whole class of bisimilar, but structurally different, models. In other words, a semantics for modal logic in the form of Kripke models has a finer structure than a semantics in terms of non-well-founded sets.
 
\cite{ger}
\end{quote}

Now, one can ask whether there exists a BK-sentence in NWF framework that can create a self-referential paradox. In order to answer this question, we will need some arguments from category theory.

\subsection{Games Played on Non-well-founded Sets}

Finally, note that the BK paradox is about the belief of the players. However, we can use NWF sets not only to represent the epistemics of games but also to represent games themselves in extensive form. As we have emphasized earlier, Aczel's graph theoretical approach identifies directed graphs and sets, and this will be our approach here as well.

Aczel's Anti-Foundation Axiom states that for each connected rooted directed graph, there corresponds a unique set \cite{acz}. Therefore, given any rooted directed graph, we can construct a game with the set that corresponds to the given graph up to the order of players.

Therefore, now, we can use \emph{any} directed graphs, not necessarily only trees, to represent games in extensive form. Clearly, the given graph may have loops that can create infinite regress, thus creating infinite games. For instance, the situations in which some players may \emph{reset} the game, or make a move that can take  the game backward are examples of such games where cyclic representation in extensive form is needed. The crucial point here is the fact that such a set/game space exists, and is unique.

\begin{thm}{\label{teo-universal}}
For every (labeled) rooted directed connected graph, there corresponds to a unique two-player NWF belief structure up to the permutation of type spaces, and the order of players.
\end{thm}

\begin{proof}
Follows directly from the Aczel's Anti-Foundation axiom.
\end{proof}

Let us illustrate the theorem with a simple example.

\begin{ex}{\label{nwf-example}}
Consider the following labeled, connected directed graph.
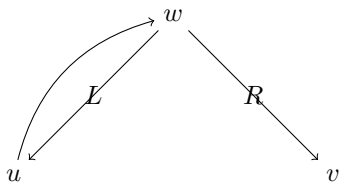
\begin{center}\begin{tikzpicture}[->,node distance=3cm]
  \node (A) {$u$};
  \node (B) [above right of=A] {$w$};
  \node (C) [below right of=B] {$v$};
  \path (B) edge node {$R$} (C)
		edge node {$L$} (A)
            (A) edge [bend left]  node {} (B);
\end{tikzpicture}\end{center}
We can now construct the two-player NWF belief structure of this game is as follows. Put $W = \{ w, v, u \}$ where $w = \{ u, v \}$ and $u = \{ w \}$. Assume that $U^a =\{w \}$, $U^b = \{ u, v\}$ (or any other combination of type spaces). Therefore, this graph corresponds to the game where Bob can \emph{reset} the game if Alice plays $L$ at $w$.
\end{ex}

In conclusion, allowing NWF sets in extensive game representations, we can express a much larger class of games. Notice that, in this section, we used NWF theory in the extensive form game representations, too - not only at belief sets.

\section{Paraconsistent Approach}

Paraconsistent logics can be captured by using several rather strong algebraic, topological and category theoretical structures. In this chapter, we  approach paraconsistency from such directions, and analyze the BK paradox within paraconsistent logics interpreted in such systems.

\subsection{Algebraic and Category Theoretical Approach}

A recent work on the BK paradox shows the general pattern of such paradoxical cases, and gives some positive results such as fixed-point theorems \cite{abr0}. In this section, we instantiate the fixed-point results of the aforementioned reference to some other mathematical structures that can represent paraconsistent logics. The surprising result, as we shall see, is the fact that even if we endorse a paraconsistent logical structure to accommodate paradoxes, there exists BK-sentences in such structures.

First, let us recall some facts about paraconsistency. Paraconsistency is the umbrella term for logical systems where \emph{ex contradictione quodlibet} fails. Namely, in paraconsistent logics, for some $\varphi, \psi$, we have $\varphi, \neg \varphi \not\vdash \psi$. 

Endorsing a paraconsistent logic does not mean that \emph{all} contradictions are true. It means that \emph{some} contradictions do not entail a trivial theory, and moreover absurdity ($\bot$) always lead to trivial theories. Thus, in paraconsistent logics, there are some contradictions which are not absurd.

Note that the semantical issue behind the failure of \emph{ex contradictione quodlibet} in paraconsistent systems is the fact that in such logical systems, the extension of the conjunction of some formulas and their negations may not be the empty set. Therefore, semantically, in paraconsistent logics, there exist some states in which a formula and its negation is true.

There are variety of different logical and algebraic structures to represent paraconsistent logics \cite{pri0}. Co-Heyting algebras are natural algebraic candidates to represent paraconsistency. We will resort to Co-Heyting algebras because of their algebraic and category theoretical properties which will be helpful later on.

\begin{dfn}
Let $L$ be a bounded distributive lattice. If there is a binary operation $\Rightarrow: L \times L \rightarrow L$ such that for all $x, y, z \in L$, $$x \leq (y \Rightarrow z) \text{ iff } (x \wedge y) \leq z,$$
then we call $(L, \Rightarrow)$ a Heyting algebra.

Dually, if we have a binary operation $\setminus: L \times L \rightarrow L$ such that $$(y \setminus z) \leq x \text{ iff } y \leq (x \vee z),$$
then we call $(L, \setminus)$ a co-Heyting algebra. We call $\Rightarrow$ implication, $\setminus$ subtraction.
\end{dfn}

An immediate example of a co-Heyting algebra is the closed subsets of a given topological space and subtopoi of a given topos \cite{law,mor,bas3,pri0}. Therefore, we can now use a closed set topology to represent paraconsistent belief sets within the BK-paradox. Such belief sets, then, would constitute a  Co-Heyting algebra. However, we need to be careful about defining the negation in such systems.

Both operations $\Rightarrow$ and $\setminus$ give rise to two different negations. The \emph{intuitionistic negation} $\dot{\neg} $ is defined as $\dot{\neg} \varphi \equiv \varphi \rightarrow \mathbf{0}$ and \emph{paraconsistent negation} ${\sim}$ is defined as ${{\sim}}\varphi \equiv \mathbf{1} \setminus \varphi$ where $\mathbf{0}$ and $\mathbf{1}$ are the bottom and the top elements of the lattice respectively. Therefore, $\dot{\neg}  \varphi$ is the largest element disjoint from $\varphi$, and ${{\sim}}\varphi$ is the smallest element whose join with $\varphi$ gives the top element $\mathbf{1}$ \cite{rey}. In a Boolean algebra both intuitionistic and paraconsistent negations coincide, and give the usual Boolean negation where  we interpret $\varphi \Rightarrow \psi$ as $\neg \varphi \vee \psi$, and $\varphi \setminus \psi$ as $\varphi \wedge \neg \psi$ with the usual Boolean negation $\neg$. What makes closed set topologies a paraconsistent structure is the fact that theories that are true at boundary points include formulas and their negation \cite{mor,bas3}. Because, a formula $\varphi$ and its paraconsistent negation ${\sim}\varphi$ intersect at the boundary of their extensions. We will discuss the paraconsistent negation in the following sections as well.

On the other hand, the algebraic structures such, as Co-Heyting algebras, we have mentioned can be approached from a category theory point of view. Before discussing Lawvere's argument, we need to define \emph{weakly point surjective} maps. An arrow $f  : A \times A \rightarrow B$ is called \emph{weakly point surjective} if for every $p : A \rightarrow B$, there is an $x : \mathbf{1} \rightarrow A$ such that for all $y : \mathbf{1} \rightarrow A$ where $\mathbf{1}$ is the terminal object, we have $$p \circ y = f \circ \langle x, y \rangle: \mathbf{1} \rightarrow B$$
In this case, we say, $p$ is represented by $x$. Moreover, a category is cartesian closed (CCC for short, henceforth), if it has a terminal object, and admits products and exponentiation. A set $X$ is said to have the fixed-point property for a function $f$, if there is an element $x \in X$ such that $f(x) = x$. Category theoretically, an object $X$ is said to have the fixed-point property if and only if for every endomorphism $f : X \rightarrow X$, there is $x : \mathbf{1} \rightarrow X$ with $ x f = x$ \cite{law0}.

\begin{thm}[\cite{law0}]{\label{law-ccc}}
In any cartesian closed category, if there exists an object $A$ and a weakly point-surjective morphism $g: A \rightarrow Y^{A}$, then $Y$ has the fixed-point property for $g$.
\end{thm}

It was observed that CCC condition can be relaxed, and Lawvere's Theorem works for categories that have only finite products \cite{abr0}\footnote{This point was already made by Lawvere and Schanuel in \emph{Conceptual Mathematics}. Thanks to Noson Yonofsky for pointing this out.}. These authors showed how to reduce Lawvere's Lemma to the BK paradox and, to reduce the BK paradox to Lawvere's Lemma\footnote{In order to be able to avoid the tecnicalities of categorical logic, we do not give the detalis of their construction and refer the reader to \cite{abr0}}. Now, our goal is to take one step further, and investigate some other cartesian closed categories which represent the non-classical frameworks that we have investigated in this paper. Therefore, by Lawvere and Abramsky \& Zvesper results, we will be able to show the existence of fixed-points in our framework, which will give the BK paradox in those frameworks. 

In their paper, Abramsky \& Zvezper, first define the BK sentence by using relations between type spaces instead of maps, and then, represent the Lawvere's fixed-point lemma in a relational framework. Then, they conclude that if the relational representation of the BK sentence satisfies some conditions then they have a fixed-point, which in turn creates the BK sentence. Their approach makes use of the standard (classical) BK paradox, and utilizes regular logic in their formalization. In our approach, we directly use Lawvere's Lemma (Theorem~\ref{law-ccc} here) to deduce our results.

Now, we observe the category theoretical properties of co-Heyting algebras and the category of hypersets. Recall that the category of Heyting algebras is a CCC. A canonical example of a Heyting algebra is the set of opens of a topological space \cite{awo0}. The objects of such a category will be the open sets. The unique morphisms in that category exists from $O$ to $O'$ if $O \subseteq O'$. What about co-Heyting algebras? It is easy to prove the following.

\begin{prop} Co-Heyting algebras are Cartesian closed categories.\end{prop}

\begin{ex}{\label{ex-coheyt}}
As we have mentioned, the co-Heyting algebra of the closed sets of a topology is a well-known example of a CCC. Similar to the arguments that show that open set topologies are CCC, we can observe that closed set topologies are CCC as well. 

Given two objects $C_{1}, C_{2}$, we define the unique arrow from $C_{1}$ to $C_{2}$, if $C_{1} \supseteq C_{2}$. The product is the union of $C_{1}$ and $C_{2}$ as the finite union of closed sets exists in a topology. The exponent $C_{1}^{C_{2}}$ is then defined as $\mathsf{Clo}(\overline{C_1} \cap C_{2})$ where $\overline{C_1}$ is the complement of $C_1$.
\end{ex}

Now, we have the following corollary for Theorem~\ref{law-ccc}.

\begin{cor}{\label{coheyt}}
In a co-Heyting algebra, if there is an object $A$ and a weakly point-surjective morphism $g: A \rightarrow Y^{A}$, then $Y$ has the fixed-point property. Therefore, there exists a co-Heyting algebraic model with an impossible BK sentence.

\end{cor}

This is interesting. In other words, even if we allow nontrivial inconsistencies and represent them as a co-Heyting algebra, we will still have fixed-points. This is our first step to establish the possibility of having the BK paradox in paraconsistent setting.

\begin{cor}
There exists a paraconsistent belief model in which the BK paradox persists.
\end{cor}

\begin{proof}
The procedure is relatively straight-forward. Take a paraconsistent belief model $M$ where belief sets constitute a closed set topology. Such a topology is a co-Heyting algebra which is a CCC. Therefore, Lawvere's Lemma (Theorem~\ref{law-ccc}) applies. Since Lawvere's Lemma is reducible to the BK sentence by Abramsky \& Zvesper 's result, we observe that $M$ possesses a fixed-point that creates the BK sentence.
\end{proof}

This simple result shows that even if we allow some contradictions, there will exists a BK sentence. This is perhaps not surprising. As Mariotti et al. pointed out earlier, interactive belief models can produce situations which are not expressible in the language \cite{mar}\footnote{In their work, however, Mariotti et al. makes the same point by resorting Cantor's diagonal arguments. Therefore, either using CCC or Cantor's diagonlization, a pattern for such self-referential paradoxes is visible \cite{yan}.}.

In our earlier discussion, we have presented some counter-models for the classical BK sentence. However, we haven't concluded that NWF models are complete. We need Lawvere's Lemma to show that NWF belief models cannot be complete. Consider the category \textbf{AFA} of hypersets with total maps between them\footnote{Thanks to Florian Lengyel for pointing this out.}. Category \textbf{AFA} admits a final object $\mathbf{1} = \{ \emptyset \}$. Moreover, it also admits exponentiation and products in the usual sense, making it a CCC. Thus, Lawvere's Lemma applies.

\begin{cor}
There exists an impossible BK sentence in non-well-founded interactive belief structures.
\end{cor}

\subsection{Topological Approach}

Now, we make it explicit how paraconsistent topological belief models are constructed. In our construction, we will make use of relational representation of belief models which in turn produces belief and assumption modalities. We will then interpret those modalities over paraconsistent topological models.

Some topological notions play an essential role in some paraconsistent logics and algebraic structures. In an early paper, Lawvere pointed out the role of boundary operator in co-Heyting algebras \cite{law}. In a similar fashion, boundaries play an essential role to give topological semantics for paraconsistent logics \cite{good,mor,bas3}. Similarly, co-Heyting algebras have been used in theories of regions and sets of regions \cite{ste2}. In the original BK paper where the paradox is first introduced, the authors discussed several complete models including topologically complete models \cite{bran0}. Their topological space, in this framework, is compact metrizable space satisfying several further conditions. Our approach, however, does not depend on the topological space \emph{per se}, but rather depends first on the logical framework we use, and then the topological space in which we interpret such a logical system.

In the topological semantics for modal logic, topological interior $\mathsf{Int}$ and closure $\mathsf{Clo}$ operators are identified with $\Box$ and $\Diamond$ operators respectively. Then, the extension $[\cdot]$ of a modal formula is given as follows $[\Box \varphi] := \mathsf{Int}([\varphi])$. In the classical setting, in general, opens or closed sets are produced by the modal operators. Thus, the extensions of Booleans may or may not be topologically open or closed. Now, we can take one step further, and stipulate that the extension of the propositional variables to be closed sets as well. Negation operator is not that straightforward in this case as the negation of a closed set is open. Therefore, we define negation ${{\sim}}$ as the \emph{closure of the complement}. Then, we obtain a co-Heyting algebra of closed sets as we have observed in Example~\ref{ex-coheyt}. In this setting, inconsistent theories are the ones that include the formulas that are true at the boundaries \cite{mor,bas3}\footnote{If we stipulate that the extension of the propositional variables to be open sets, we obtain intuitionistic logic with intuitionistic negation. Thus, the duality of intuitionistic and paraconsistent logics is rather clear in the topological semantics.}.

We now step by step construct the BK argument in paraconsistent topological setting. We will call such a belief model a \emph{paraconsistent topological belief model}.

For the agents $a$ and $b$, we have a corresponding non-empty type space $A$ and $B$, and define closed set topologies $\tau_A$ and $\tau_B$ on $A$ and $B$ respectively. Furthermore, in order to establish connection between $\tau_A$ and $\tau_B$ to represent belief interaction among the players, we introduce additional constructions $t_A \subseteq A \times B$, and  $t_B \subseteq B \times A$. We then call the structure $F = (A, B, \tau_A, \tau_B, t_A, t_B)$ a paraconsistent topological belief model. In this setting, the set $A$ represents the possible epistemic states of the player $a$ in which she holds beliefs about player $b$, or about $b$'s beliefs etc, and vice versa for the set $B$ and the player $a$. Moreover, the topologies represent those beliefs. For instance, for player $a$ at the state $x \in A$, $t_A$ returns a closed set in $Y \in \tau_B \subseteq \wp{(B)}$. In this case, we write $t_A(x, Y)$ which means that at state $x$, player $a$ believes that the states $y$ in $Y \in \tau_B$ are possible for the player $b$, i.e. $t_A(x, y)$ for all $y \in Y$. Moreover, a state $x \in A$ \emph{believes} $\varphi \subseteq B$ if $\{ y : t_A(x, y) \} \subseteq \varphi$. Furthermore, a state $x \in A$ \emph{assumes} $\varphi$ if $\{ y : t_A(x, y) \} = \varphi$. Notice that in this definition, we identify logical formulas with their extensions.

The modal language which we use has two modalities representing the beliefs of each agent. Akin to some earlier modal semantics for the paradox, we give a topological semantics for the BK argument in paraconsistent topological belief models \cite{bran0,pac2}. Let us first give the formal language which we use. The language for the logic topological belief models is given as follows.
$$\varphi : = p ~|~ {{\sim}} \varphi ~|~ \varphi \wedge \varphi ~|~ \Box_a ~|~ \Box_b ~|~ \boxplus_a ~|~ \boxplus_b$$
where $p$ is a propositional variable, ${{\sim}}$ is the paraconsistent topological negation symbol which we have defined earlier, and $\Box_i$ and $\boxplus_i$ are the belief and assumption operators for player $i$, respectively.

We have discussed the semantics of the negation already. For $x \in A$, $y \in B$, the semantics of the modalities are given as follows with a modal valuation attached to $F$. 
\begin{quote}\begin{tabular}{llll}
$x \models \Box_a \varphi$ & iff & $\exists Y \in \tau_B$ with $t_A(x, Y) \rightarrow \forall y \in Y . y \models \varphi$\\
$x \models \boxplus_a \varphi$ & iff & $\exists Y \in \tau_B$ with $t_A(x, Y) \leftrightarrow \forall y \in Y .  y \models \varphi$\\
$y \models \Box_b \varphi$ & iff & $\exists X \in \tau_A$ with $t_B(y, X) \rightarrow \forall x \in X . x \models \varphi$\\
$y \models \boxplus_b \varphi$ & iff & $\exists X \in \tau_A$ with $t_B(y, X) \leftrightarrow \forall x \in X . x \models \varphi$
\end{tabular}\end{quote}
We define the dual modalities $\Diamond_a$ and $\Diamond_b$ as usual. 

Now, we have sufficient tools to represent the BK sentence in our paraconsistent topological belief structure with respect to a state $x_0$: $$x_0 \models \Box_a \boxplus_b \varphi \wedge \Diamond_a \top$$

Let us analyze this formula in our structure. Notice that the second conjunct guarantees that for the given $x_0 \in A$, there exists a corresponding set $Y \in \tau_B$ with $t_A(x, Y)$. On the other hand, the first conjunct deserves closer attention:
\begin{quote}\begin{tabular}{lll}
$x_0 \models  \Box_a \boxplus_b \varphi$ & iff & $\exists Y \in \tau_B$ with $t_A(x_0, Y) \Rightarrow \forall y \in Y .~ y \models \boxplus_b \varphi$ \\
& iff & $\exists Y \in \tau_B$ with $t_A(x_0, Y) \Rightarrow$ \\
&& $[ \forall y \in Y, \exists X \in \tau_A \text{ with } t_B(y, X) \Leftrightarrow \forall x \in X .~ x \models \varphi ]$
\end{tabular}\end{quote}

Notice that in our framework, some special $x$ can satisfy falsehood $\bot$ to give $x_0 \models  \Box_a \boxplus_b \bot \wedge \Diamond_a \top$ for some $x_{0}$. Let the extension of $p$ be $X_{0}$. Pick $x_{0} \in \partial X_{0}$ where $\partial(\cdot)$ operator denotes the boundary of a set $\partial(\cdot) = \mathsf{Clo}(\cdot) - \mathsf{Int}(\cdot)$. By the assumptions of our framework $X_{0}$ is closed. Moreover, by simple topology $\partial X_{0}$ is closed as well. By the second conjunct of the formula in question, we know that some $Y \in \tau_{B}$ exists such that $t_{A}(x_{0}, Y)$. Now, for all $y$ in $Y$, we make an additional supposition and associate $y$ with $\partial X_{0}$ giving $t_{B}(y, \partial X_{0})$. We know that for all $x \in \partial X_{0}$, we have $x \models p$ as $\partial X_{0} \subseteq X_{0}$ where $X_{0}$ is the extension of $p$. Moreover, $x \models {{\sim}} p$ for all $x \in \partial X_{0}$ as $\partial X_{0} \subseteq ({{\sim}} X_{0})$, too. Thus,  we conclude that $x_0 \models  \Box_a \boxplus_b \bot \wedge \Diamond_a \top$ for some \emph{carefully selected} $x_{0}$.

In this construction, we have several suppositions. First, we picked the actual  state from the boundary of the extension of some proposition (ground or modal). Second, we associate the epistemic accessibility of the second player to the same boundary set. Namely, $a$'s beliefs about $b$ includes her current state.

Now, the BK paradox appears when one substitutes $\varphi$ with the following diagonal formula (whose extension is a closed set by definition of the closed set topology), hence breaking the aforementioned circularity: $$D(x) = \forall y. [t_A(x, y) \rightarrow~ {{\sim}} t_B (y, x)]$$ The BK impossibility theorem asserts that, under the seriality condition, there is no such $x_0$ satisfying the following.
\begin{quote}\begin{tabular}{lll}
$x_0 \models  \Box_a \boxplus_b D(x)$ & iff & $\exists Y \in \tau_B$ with $t_A(x_0, Y) \Rightarrow$ \\
& & $[ \forall y \in Y, \exists X \in \tau_A \text{ with } t_B(y, X) \Leftrightarrow$ \\
&& $\forall x \in X .~ x \models \forall y'. (t_A(x, y') \rightarrow~ {{\sim}} t_B (y', x)) ]$
\end{tabular}\end{quote}

Motivated by our earlier discussion, let us analyze the logical statement in question. Let $X_0$ satisfy the statement $t_A(x, y')$ for all $y' \in Y$ and $x \in X_0$ for some $Y$. Then, $\partial X_0 \subseteq X_0$ will satisfy the same formula. Similarly, let ${{\sim}} X_0$ satisfy ${{\sim}} t_B(y', x)$ for all $y' \in Y$ and $x \in X_0$. Then, by the similar argument, $\partial ({\sim} X_0)$ satisfy the same formula. Since $\partial(X_0) = \partial({\sim} X_0)$, we observe that any $x_0 \in \partial X_0$ satisfy $t_A(x, y')$ and ${{\sim}}t_B(y', x)$ with the aforementioned quantification. Thus, such an $x_0$ satisfies $\Box_a \boxplus_b D(x)$. Therefore, the states at the boundary of some closed set \emph{satisfy} the BK sentence in paraconsistent topological belief structures. Thus, this is a counter-model for the BK sentence in the paraconsistent topological belief models.

\begin{thm}The BK sentence is satisfiable in some paraconsistent topological belief models.\end{thm}

\begin{proof} See the above discussion for the proof which gives a model that satisfies the BK sentence. \end{proof}

\subsection{Product Topologies}

In the previous section, we introduced $t_A$ and $t_B$ to represent the belief interaction between the players. However, topological models provide us with some further tools to represent such an interaction \cite{gabb}.

In this section, we use product topologies to represent belief interaction among the players. Novelty of this approach is not only to economize on the notation and the model, but also to present a more natural way to represent the belief interaction. For our purposes here, we will only consider two-player games, and our results can easily be generalized to $n$-player. Here, we make use of the constructions presented in some recent works \cite{ben1,ben3}.

\begin{dfn}
Let $a, b$ be two players with corresponding type space $A$ and $B$. Let $\tau_A$ and $\tau_B$ be the (paraconsistent) closed set topologies of respective type spaces. The product topological paraconsistent belief structure for two agents is given as $(A \times B, \tau_A \times \tau_B)$.
\end{dfn}
 
In this framework, we assume that the topologies are \emph{full} on their sets - namely $\bigcup \tau_A = A$, and likewise for $B$. In other words, we do not want any non-expressibility results just because the given topologies do not cover such states. If the topologies are not full, we can reduce the given space to a subset of it on which the topologies are full.

In this setting, if player $a$ believes proposition $P \subseteq B$ at state $x \in A$, we stipulate that there is a closed set $X \in \tau_A$ such that $x \in X$ and a closed set $Y \in \tau_B$ with $Y \subseteq P$, all implying $X \times Y \in \tau_A \times \tau_B$. Player $a$ assumes $P$ if $Y = P$, and likewise for player $b$. Similar to the previous section, we make use of paraconsistent topological structures with closed sets and paraconsistent negation.

Borrowing a standard definitions from topology, we say that given a set $S \subseteq A \times B$, we say that $S$ is \emph{horizontally closed} if for any $(x, y) \in S$, there exists a closed set $X$ with $x \in X \in \tau_A$ and $X \times \{ y \} \subseteq S$. Similarly, $S$ is \emph{vertically closed} if for any $(x, y) \in S$, there exists a closed set $Y$ with $y \in Y \in \tau_B$, and $\{ x \} \times Y \subseteq S$  \cite{ben1,ben3}. In this framework, we say player $a$ at $x \in A$ is said to believe a set $Y \subseteq B$ if $\{ x \} \times Y$ is vertically closed. 

Now, we define assumption-complete structures in product topologies. For a given language $\mathcal{L}$ for our belief model, let $\mathcal{L}^a$ and $\mathcal{L}^b$ be the families of all subsets of $A$ and $B$ respectively. Then, we observe that by assumption-completeness, we require every non-empty set $Y \in \mathcal{L}^b$ is assumed by some $x \in A$, and similarly, every non-empty set $X \in \mathcal{L}^a$ is assumed by some $y \in B$.
 
We can now characterize assumption-complete paraconsistent topological belief models. Given type spaces $A$ and $B$, we construct the coarsest topologies on respective type spaces $\tau_A$ and $\tau_B$ where each subset of $A$ and $B$ are in $\tau_A$ and $\tau_B$. Therefore, it is easy to see that $A \times B$ is vertically and horizontally closed for any $S \subseteq A \times B$. Moreover, under these conditions, our belief structure in question is assumption-complete. 

We can relax some these conditions. Assume that now $\tau_A$ and $\tau_B$ are not the coarsest topologies on $A$ and $B$ respectively. Therefore, we define, we \emph{weak assumption-completeness} for a topological belief structure if every set $S \in A \times B$ is both horizontally and vertically closed. In other words, weak assumption-complete models focus only on the formula that are available in the given structure. There can be some formulas expressible in $\mathcal{L}$, but not available in $\mathcal{L}^a$ or in $\tau_A$ for some reasons. Epistemic game theory, indeed, is full of such cases where players may or may not be allowed to send some particular signals, and some information may be unavailable to some certain players. The following theorem follows directly from the definitions.

\begin{thm}
Let $M = (A \times B, \tau_A \times \tau_B)$ be a product topological paraconsistent belief model. If $M$ is horizontally and vertically closed, then it is weak assumption-complete.
\end{thm}

\section{Conclusion and Future Work}

In this paper, we presented two non-classical frameworks to formalize beliefs in games. Non-well-founded sets, among many other things, enabled us to use a larger collection of graphs to represent circular games where some players may \emph{reset} the games, and possibly loop around themselves\footnote{Under the absence of rationality, perhaps.}. Paraconsistent logic, on the other hand, gave us the tools to represent inconsistent beliefs of the players.

Paraconsistency presents a rather interesting framework for game theory \cite{rah}. Therefore, a possible next step in this direction would be to present a paraconsistent epistemic game theory where the \emph{knowledge sets} of the agents may be inconsistent. Our work so far has lead us to consider paraconsistent belief sets in games where the belief sets of players may contain contradictions without leading to triviality so that we can still make meaningful inferences\footnote{Thanks to Graham Priest for drawing my attention to this point.}. Then, the question is the following: ``What does it mean to make an inconsistent move?" Dialogical games seems to provide some answers to this question \cite{rah}. 

Moreover, from a logical perspective, paraconsistency has its dual intuitionistic form where the belief sets of the agents maybe \emph{incomplete} or \emph{paracomplete}. Therefore, for some player $i$, there can be a formula $\varphi$ for which neither $\varphi$ nor its negation is believed by any player. This is an intriguing aspect for game where players simply do not know what to play. The concept of \emph{empty move} may seem meaningful for such issues.

Our work raises the question of paraconsistent games. In this work, we focused on the paraconsistent belief sets, but one can very well start with perfect information games with non-classical probabilities. In such cases, sum of the probabilities of playing $a$ and not-$a$ may be higher than 1\cite{wil0}\footnote{Thanks to Brian Weatherson for pointing this out.}. 

Therefore, we conclude that aforementioned non-classical frameworks provide a richer framework to analyze issues in foundational and epistemic game theory, and further work on the subject will shed some light on the \emph{paradoxes} within the field.

\paragraph{Acknowledgements} Thanks to Florian Lengyel, Rohit Parikh, Graham Priest, Noson Yonofsky.

\bibliographystyle{authordate3}  
\bibliography{/Users/can/Documents/Akademik/papers.bib} 
\end{document}